\documentclass[]{spie}
\usepackage[]{graphicx}
\usepackage{amsmath}
\usepackage{enumerate}

\title{The nonlinear photon transfer curve of CCDs \\
and its effects on photometry}

\author{Bin Ma\supit{1}, Zhaohui Shang\supit{2,1}, Lifan Wang\supit{3,4}, Yi Hu\supit{1}, Qiang Liu\supit{1}, Peng Wei\supit{1}
\skiplinehalf
\supit{1}National Astronomical Observatories, Chinese Academy of Sciences, Beijing, China; \\
\supit{2}Tianjin Normal University, Tianjin, China; \\
\supit{3}Purple Mountain Observatory, Chinese Academy of Sciences, Nanjing, China; \\
\supit{4}Texas A\&M University, College Station, Texas, USA;
}

\authorinfo{Send correspondence to Bin Ma: mabin22@gmail.com}

\begin{document}
  \maketitle

\begin{abstract}

The photon transfer curve (PTC, variance vs. signal level) is a
commonly used and effective tool in characterizing CCD 
performance. It is theoretically linear in the range where photon shot
noise dominates, and its slope is utilized to derive the gain of the
CCD. However,
recent researches on different CCDs have revealed that the variance
progressively drops at high signal levels, while the linearity shown
by signal versus
exposure time is still excellent and unaffected. On the other hand,
bright stars are found to exhibit fatter point spread function (PSF).
Both nonlinear PTC and the brighter-fatter effect are regarded as the
result of spreading of charges between pixels, an interaction progress
increasing with signal level. In this work we investigate the
nonlinear PTC based on the images with a STA1600FT CCD camera, whose
PTC starts to become nonlinear at about 1/3 full well. To explain the
phenomenon, we present a model to characterize the charge-sharing PSF.
This signal-dependent PSF can be derived from flat-field frames, and allow
us to quantify the effects on photometry and measured shape of stars.  This
effect is essentially critical for projects requiring accurate
photometry and shape parameters.

\end{abstract}


\section{INTRODUCTION}
\label{sec:intro}

The photon transfer curve (PTC, variance vs. signal
level)\cite{Janesick2000} is a fundamental tool in characterizing
Charge-Coupled Device (CCD) performance. Its primary purpose is to
determine the ``gain'' parameter of a CCD system which denotes the conversion from the
number of collected electrons to the digital number read out by the amplifier.
Using the gain, many of the other performance parameters such as
readout noise, dark current, full well etc.\ are converted in physical
unit of $e^-$. 

The variance of pixel values in a flat frame with uniform illumination
consists of: 

\begin{enumerate}[i)] 

\item readout noise, which is constant and usually small in modern 
CCDs;

\item photon shot noise, which follows Poisson noise and is proportional
to the signal level;

\item actual signal fluctuation between pixels, which is caused by
the Quantum Efficiency (QE) difference and is
proportional to the square of the signal level.

\end{enumerate} 

If a pair of flat frames are taken with identical illumination and
exposure parameters, their
difference will remove the QE-related sensitivity fluctuation. Therefore the
common PTC is linear beyond extremely low signal, and its slope is the
inverse of the ``gain'' parameter. 

However, Ref.\,\citenum{Downing2006} reported that the nonlinearity of
PTC had been observed on CCDs from different manufacturers even though
the signal increased excellently linearly with exposure time ( better
than $1\%$ ). In addition, the PTC nonlinearity progressively
increased with signal level, and was in excess of $15\%$ before
saturation was reached. They carried out many tests and excluded some
potential causes such as lateral diffusion of charge in the undepleted
region at the back of the imager, the clocking or transport of charge
in the image area or serial register to the read out, the output
amplifier, and the detector electronics. They concluded that the nonlinear
PTC was related to the amount of charge collected within a pixel, and
suggested that the mechanism behind nonlinearity was charge spreading
or sharing according to spatial autocorrelation analysis. Their other
work \cite{Downing2009} shows the nonlinearity is greater for thicker
devices made from higher resistivity silicon. 

To explain the  phenomenon, they have presented a
hypothesis\cite{Downing2013} that the collected charges in a pixel
reduces the extend of the pixel's electric field, and consequently
reduces the pixel's collection competitiveness with respect to its
neighbors. Because the charge sharing increases linearly with signal,
they expected the Point Spread Function (PSF) to broaden with signal,
and indeed found that the PSF of laboratory spots rose linearly with
their signal levels. This brighter-fatter effect was detected in scientific
images as well\cite{Astier2013}. In addition, they suggested that the PTC
could be well fitted to a quadratic function, the linear term of
which should be a better estimator of gain.

Alternatively, Ref.\,\citenum{Stefanov2014} developed a Monte Carlo model
to describe the process of signal dependent charge-sharing. They
formulated the probability of charge-sharing to be proportional to the
instantaneous number of excess electrons in a pixel compared to the
average number of electrons over all pixels. And the shared electrons could be
received by either a random neighbor pixel or the neighbor with the lowest
number of electrons. Then a Monte Carlo simulation, tracing
individual electrons generated by Poisson-distributed photons,
reconstructs the quadratic PTC with sub-Poisson noise. However, the
model is a macro model, of which the sharing coefficient is empirical
but not based on physical parameters.

In another model by Ref.\,\citenum{Antilogus2014}, the Coulomb forces
from accumulated charges within all pixels of the CCD distort the drift electric
field, making the boundaries of each pixel slightly dynamical and hence
forthcoming charges probably drift on the boundaries. This model
bridges both the lab tests and scientific observations: the
displacement of the effective boundaries of a pixel at some signal
level is derived from correlations in the lab flat frames, then it is
applied in the scientific images and predicts the flux-dependent PSF
with $\sim 20\%$ accuracy. However, for a $n\times n$ region in
question, there need $4n^2$ coefficients for
a pixel with four boundaries and $n^2$ neighbors, while there are only
$n^2$ correlation measurements. Providing that each boundary is shared
by two pixels, the measurements are still short by a fact of $2$. In
order to solve the problem anyway, they have imposed ratios of coefficients
addressing similar source-test separations.

We have found that the CCDs for the trio Antarctic Survey Telescopes
(AST3)\cite{Cui2008} also exhibit signal dependent nonlinear
PTC\cite{Ma2012}. The CCD cameras, STA1600FT, are designed and
manufactured by Semiconductor Technology Associates, Inc.. They have
$10,560 \times 10,560$ pixels with a pixel size of 9 $\mu$m, corresponding
to 1 arcsec on the focal plane of AST3. The CCDs work in frame transfer
mode to avoid malfunction of mechanical shutter, and they have $16$
amplifiers to speedup the readout. In this paper, we present an
alternative model of charge sharing to study the nonlinear PTC. We
will investigate the PTC and correlation in the lab flat frames in
Sec.~\ref{sec:ptc}. In Sec.~\ref{sec:model} we describe the model,
which simply takes into account the final integral effect of 
charge-sharing depending on signal level. With the charge-sharing PSF results from flat
frames, we discuss the effects on photometry in
Sec.~\ref{sec:effect}, and finally conclude in
Sec.~\ref{sec:conclusion}.

\section{PTC NONLINEARITY AND AUTOCORRELATION}
\label{sec:ptc}

\begin{figure}
   \begin{center}
   \begin{tabular}{c}
   \includegraphics[height=7cm]{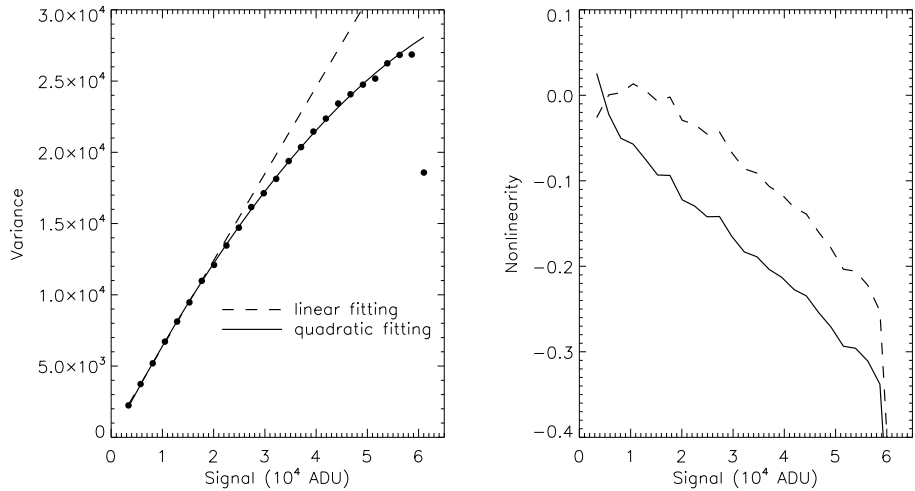}
   \end{tabular}
   \end{center}
   \caption[ptc]
   {\label{fig:ptc} PTC (left) and its nonlinearity (right) of AST3
CCD. In the left panel, the filled circles are data points, and the
variance tends to dip as the signal level increases. The sudden drop
of variance at the highest signal level is due to saturation. The
dashed line is the best linear fitting for $signal < 20,000$, and the
solid line is the best quadratic fitting to all data points except the
saturated one. The right panel shows the normalized residuals with
regard to the linear
fitting (dashed line) and to only the linear part of the quadratic fitting (solid
line). } \end{figure}

We investigate the PTC of AST3 CCD based on the lab flat frames. The
PTC is plotted in Fig.~\ref{fig:ptc}, and it becomes nonlinear
starting 
from  $signal \sim 20,000$ ADU. Therefore in Ref.\,\citenum{Ma2012} we
fitted the data points with $signal  < 20,000$ ADU with a straight line, whose
slope is the inverse of gain. The slope of the best linear fitting in
Fig.~\ref{fig:ptc} is $0.61$, corresponding to a $gain = 1.64$
$e^-$/ADU. The PTC tends to deviate from the linear fitting from
$signal \sim 20,000$ ADU, and the nonlinearity exceeds $20\%$ at
$signal \sim 60,000$ ADU.

Alternatively a quadratic polynomial has been proved to be an
excellent fit to the entire signal range \cite{Downing2013,
Stefanov2014}, and the coefficient of its linear term can be used to
derive the gain. Fig.~\ref{fig:ptc} illustrates that the quadratic
function also fits the data of AST3 CCD quite well up to the saturated
signal level. The coefficient of its linear term of $0.69$ results in $gain =
1.44$ $e^-$/ADU, which is $12\%$ lower than the previous value from
linear fitting to lower signal data. Besides, the nonlinearity occurs
obviously even when the signal is lower, and rises to $\sim 30\%$ at
$signal \sim 60,000$ ADU.

The sub-Poisson noise strongly suggests that neighboring pixels of
flat-fields have statistical correlations\cite{Downing2006,
Downing2013, Antilogus2014}. If $D_{i,j}$ ($i,j$ are the pixel
coordinates) is the difference of two
uniformly illuminated flat frames, the 2D spatial autocorrelation
function $R_{m,n}$, which denotes the correlation between $D_{i,j}$
and $D_{i+m,j+n}$, is calculated by:

	\begin{equation}
	\label{eq:autocorr}
R_{m,n} = \frac{\sum\limits_{i=1,j=1}^{i=M-m,j=N-n} D_{i,j}D_{i+m,j+n}}{\sum\limits_{i=1,j=1}^{i=M-m,j=N-n}D_{i,j}^2} .
	\end{equation}

\begin{figure}
   \begin{center}
   \begin{tabular}{c}
   \includegraphics[height=11cm]{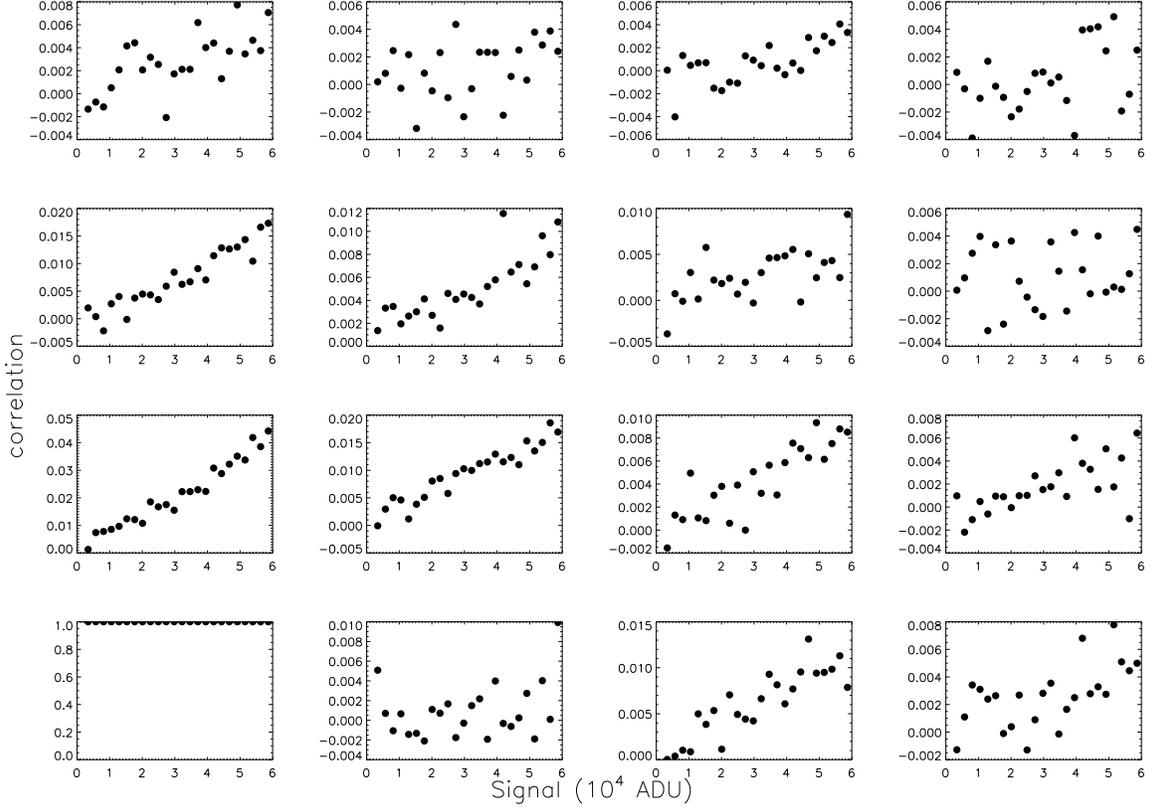}
   \end{tabular}
   \end{center}
   \caption[corr]
   {\label{fig:corr} The correlation coefficients $R_{m,n}$ grid for
$0 \leq m \leq 3$, $0 \leq n \leq 3$ as a function of signal level.}
   \end{figure}

For the ideal Poisson noise, each pixel is independent, thus $R_{m,n}
= 0$ (except $R_{0,0} \equiv 1$). Due to the parity, we only calculate
one quadrant of $R_{m,n}$ i.e. $m \geq 0, n \geq 0$, and plot the
signal dependent $R_{m,n}$ with only a $4 \times 4$ grid ($0 \leq m \leq 3,
0 \leq n \leq 3$) in Fig.~\ref{fig:corr}. The correlations rise with
signal level, decay rapidly with separation, and are larger along
Y-axis (CCD columns, vertical transfer) than along X-axis. The nearest vertical neighbors
exhibit the strongest correlation, reaching up to about $4.5\%$ at its
highest signal level.  
The second nearest vertical neighbors and the
diagonal neighbors show the second strongest correlation, reaching to
about $2\%$.  However, the nearest horizontal neighbors
exhibit nearly negligible correlation, whereas the second nearest
horizontal neighbors show a merely correlation with a maximum of $\sim
1\%$. 
For farther distances, there seems to be weaker or negligible
correlations.
We believe the positive correlations seen here indicate that the
nonlinear PTC is indeed caused by the charge-sharing effect.

\section{CHARGE-SHARING PSF MODEL}
\label{sec:model}

To explain the nonlinearity of PTC and the correlations between pixels, we
present a semi-quantitative model on how many charges generated in one pixel are
shared with its neighbors. Compared to the model of
Ref.\,\citenum{Stefanov2014}, which takes into account the individual
electrons and the model of Ref.\,\citenum{Antilogus2014}, which
estimates the pixel boundary displacements perturbed by each
neighbor, our model directly derives the final charge-sharing
fractions into its neighbors, which is a function of the distance of
the neighbors.  This function virtually results in an extra PSF, which 
we call the charge-sharing PSF.
Because the Coulomb forces from stored electrons decrease the
potential wells of pixels, the subsequently generated electrons have some
probability to diffuse and may be collected by the neighbor pixels.
The shared fraction increases as charges are accumulated, therefore
the charge-sharing PSF also depends on the signal level, different from
the pure space-varying PSF usually caused by the optics of a telescope and atmospheric seeing.


Deriving the charge-sharing PSF will allow us to quantitatively
characterize the CCD's performance as well as to make potential
corrections to astronomical images to reduce uncertainties in
photometry.

We begin with denoting the charge-sharing PSF as $p_{i,j}$, $-n \leq i,j \leq n$,
where $n$ is the size of the PSF, and $p_{i,j} \geq 0, \sum p_{i,j} \equiv
1$. For a certain pixel, its signal without charge sharing is
$I_{0,0}$, then after charge sharing when some charges are lost and
some extra charges are received, the signal becomes $I_{0,0}'$:

	\begin{equation}
	\label{eq:signal}
I_{0,0}' = \sum\limits_{i=-n,j=-n}^{i=n,j=n} p_{i,j} I_{-i,-j} .
	\end{equation}
Then its variance $V'$ for the uniformly illuminated flat frame as a function of raw variance $V$ is:
	\begin{equation}
	\label{eq:variance}
V' = \sum\limits_{i=-n,j=-n}^{i=n,j=n} p_{i,j}^2 V .
	\end{equation}

Since $p_{i,j} \geq 0, \sum p_{i,j} \equiv 1$, we have $V' < V$ unless
$p_{0,0} = 1$ which means no charge sharing. So the charge sharing
smoothes the image and reduces the variance, as seen in PTC.

The correlation $R_{m,n}$ between $I_{0,0}'$ and $I_{m,n}'$ can be
expressed from charge-sharing PSF by:

	\begin{equation}
	\label{eq:correlation}
R_{m,n} = \frac{Cov(I_{0,0}',I_{m,n}')}{V'} = \frac{Cov(\sum\limits_{i,j} p_{i,j} I_{-i,-j}, \sum\limits_{i,j} p_{i,j} I_{m-i,n-j})}{\sum p_{i,j}^2 V} ,
	\end{equation} 
where $Cov$ denotes the covariance, and $Cov(I_{i,j},I_{m,n}) = 0$ unless $i=m, j=n$ when $Cov(I_{i,j},I_{m,n}) = V$. Therefore the correlations arise due to the shared charges from the same pixels:

	\begin{equation}
	\label{eq:correlation2}
R_{m,n} = \frac{\sum\limits_{i,j} p_{-i,-j} p_{m-i,n-j}}{\sum p_{i,j}^2} .
	\end{equation}  

Approximately, we only consider the terms showing significant
correlation in Sec.~\ref{sec:ptc} (see Fig.~\ref{fig:corr}), which are $p_{0,0}$, $p_{0,\pm 1}$,
$p_{0,\pm 2}$, and $p_{\pm 1, \pm 1}$, respectively. Hence
Eq.~\ref{eq:correlation2} can be written as:

	\begin{equation}
	\label{eq:correlation3}
	\begin{split}
R_{0,1} &= \frac{2p_{0,0}p_{0,1} + 2p_{0,1}p_{0,2}}{\sum p_{i,j}^2} \\
R_{0,2} &= \frac{2p_{0,1}^2 + 2p_{0,0}p_{0,2}}{\sum p_{i,j}^2} \\
R_{1,1} &= \frac{2p_{0,0}p_{1,1} + 2p_{1,1}p_{0,2}}{\sum p_{i,j}^2} 
	\end{split}
	\end{equation} 

To solve these PSF coefficients from the PTC and correlation
measurements, we take further approximations. Since $p_{0,0}$ is the
largest and close to $1$, $p_{0,0}^2$ term dominates the variance in
Eq.~\ref{eq:variance}, consequently the remained fraction $p_{0,0}$
can be estimated by:

	\begin{equation}
	\label{eq:p00}
p_{0,0} \simeq \sqrt{V' / V} .
	\end{equation}
Also allowing for $p_{0,0} \gg p_{0,1} \sim p_{0,2} \sim p_{1,1}$, the sharing fractions to neighbor pixels in Eq.~\ref{eq:correlation3} are approximated to:
	\begin{equation}
	\label{eq:p01p11}
	\begin{split}
p_{0,1} &\simeq p_{0,0} R_{0,1} / 2 \\
p_{0,2} &\simeq p_{0,0} R_{0,2} / 2 \\
p_{1,1} &\simeq p_{0,0} R_{1,1} / 2 . 
	\end{split}
	\end{equation} 
It is clear that we can calculate the charge-sharing PSF $p_{i,j}$ from the
data we have in Fig.~\ref{fig:ptc} and Fig.~\ref{fig:corr}.

   \begin{figure}
   \begin{center}
   \begin{tabular}{c}
   \includegraphics[height=9cm]{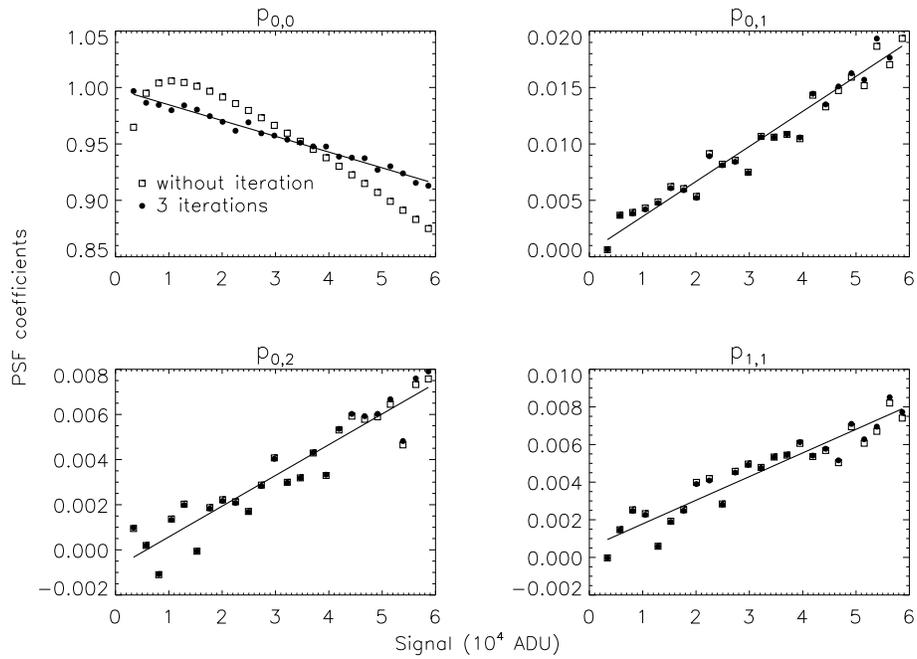}
   \end{tabular}
   \end{center}
   \caption[psflin]
   {\label{fig:psflin} The signal dependent charge-sharing PSF 
coefficients from the linear fitting (Fig.~\ref{fig:ptc}).  Open
squares are initial estimates 
and the filled circles are the convergent values after 3 
iterations.}
   \end{figure}

   \begin{figure}
   \begin{center}
   \begin{tabular}{c}
   \includegraphics[height=9cm]{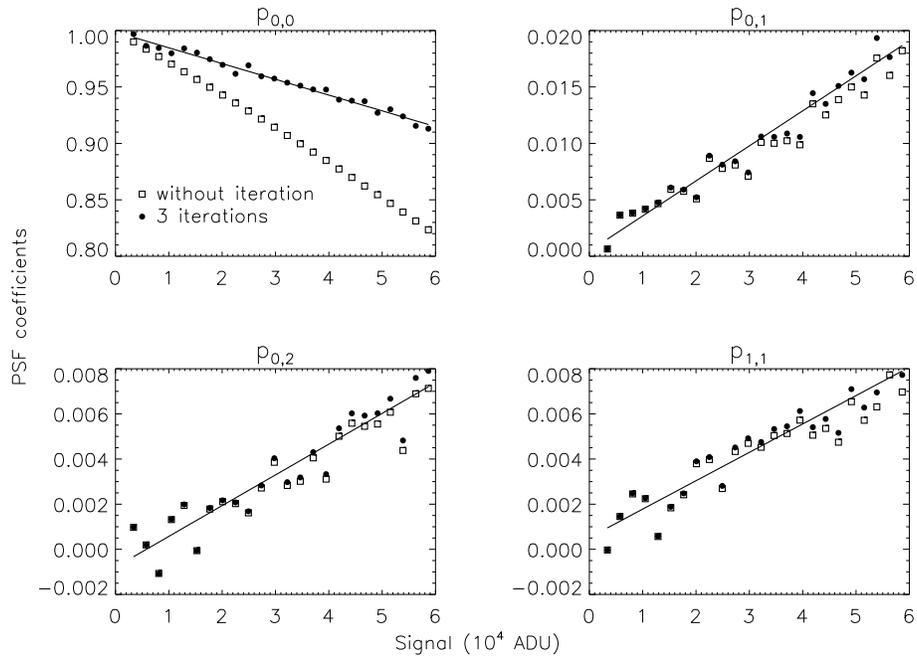}
   \end{tabular}
   \end{center}
   \caption[psfqua]
   {\label{fig:psfqua} Same as Fig.~\ref{fig:psflin}, but for the
quadratic fitting to the entire PTC (Fig.~\ref{fig:ptc}).}
   \end{figure}

   \begin{figure}
   \begin{center}
   \begin{tabular}{l}
   \includegraphics[height=6cm]{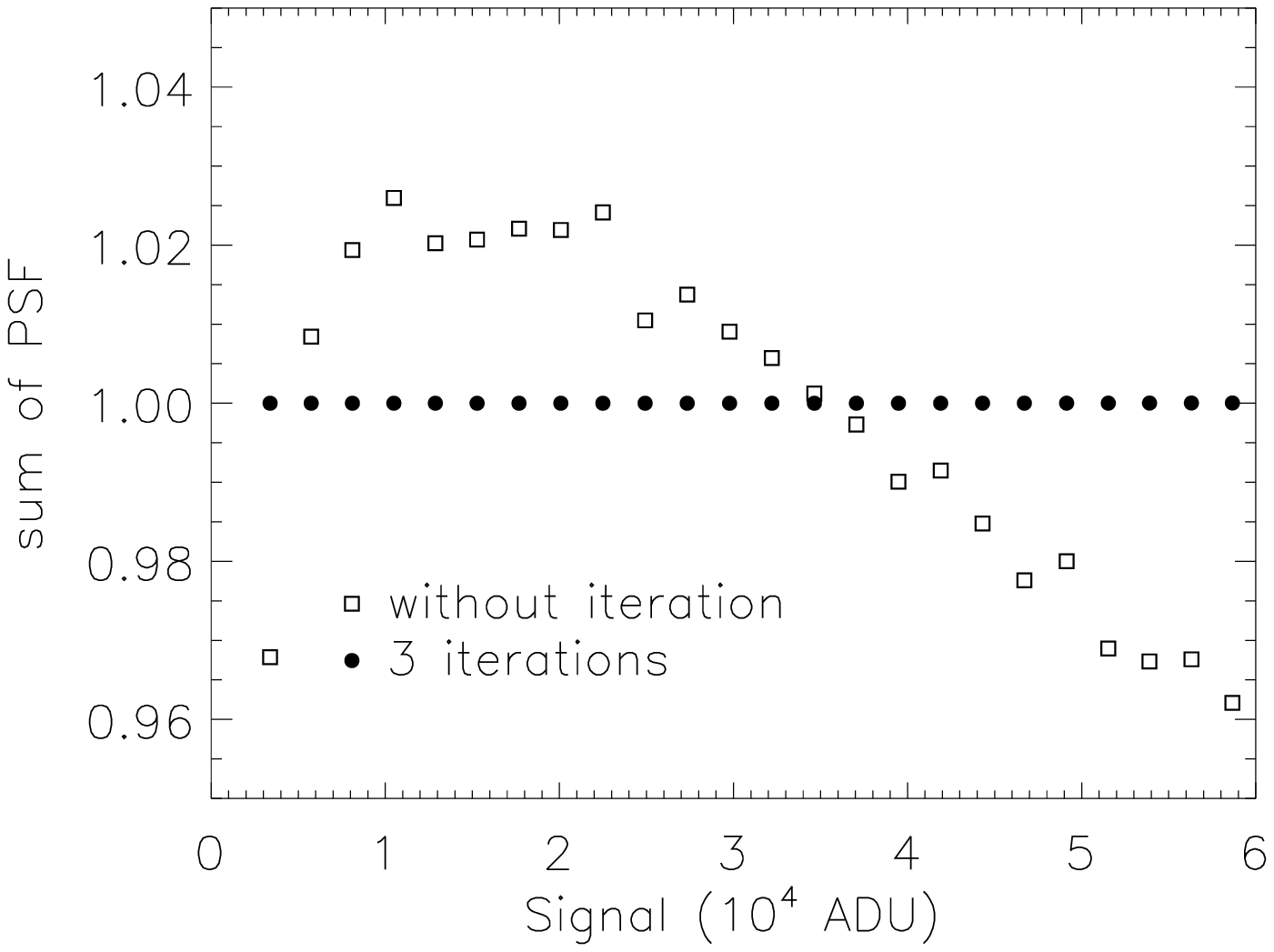}
   \includegraphics[height=6cm]{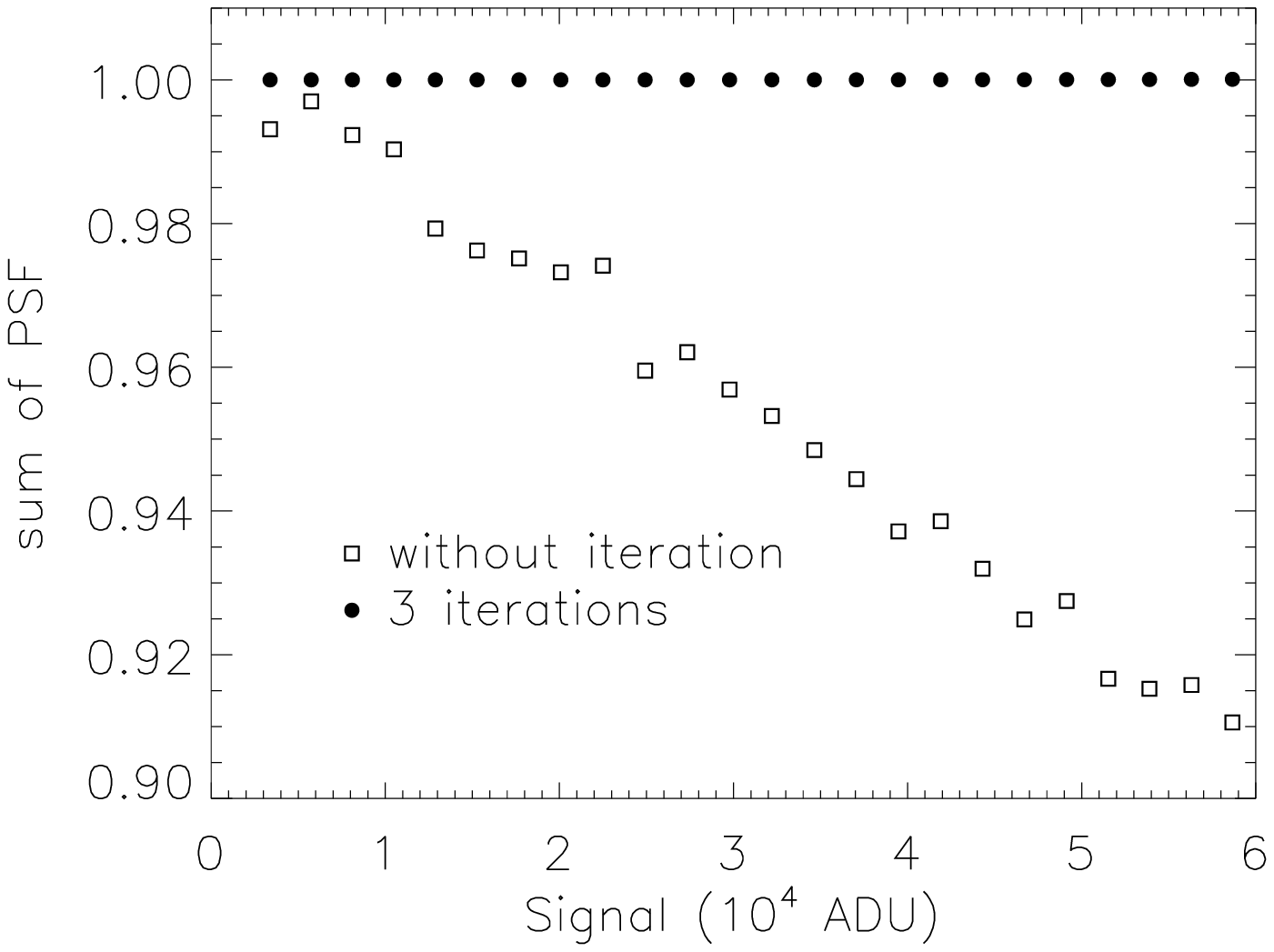}
   \end{tabular}
   \end{center}
   \caption[sumpsf]
   {\label{fig:sumpsf} The sum of the charge-sharing PSF coefficients, which is expected
to be exact $1$. Open squares represent the initial estimations where
the ideal Poisson variances $V$ are estimated from PTC
(Fig.~\ref{fig:ptc}) either by
linear fitting (left) or by the linear term of the quadratic
fitting (right). The filled circles
represent the results after three iterations where the ideal
Poisson variances are calculated by the latest derived PSF.}
   \end{figure}

However the ideal Poisson variances $V$ estimated by linear fitting to
low signal and by the linear term of the quadratic fitting to the entire signal
range (Sec~\ref{sec:ptc}) vary by a factor of $\sim 10\%$. We utilize both linear and
quadratic fittings to calculate $p_{0,0}$ and subsequent coefficients,
and plot them as open squares in Fig.~\ref{fig:psflin} and
Fig.~\ref{fig:psfqua}, respectively. As it has a slightly less
nonlinearity, the linear fitting (initial estimates) results in an apparently larger
$p_{0,0}$. But both methods end up with shrinking sums of PSF when
the signal increases (see Fig.~\ref{fig:sumpsf}), despite that the
sums are expected to be invariably unity. Moreover, the sums of PSF from
the quadratic fitting suffer greater deviation, whose maximum is near
$10\%$ at the highest signal level, while the maximal deviation from
the linear fitting is less than $4\%$. 
 
To reduce the uncertainties, we employ an iteration method. We re-derive
the remained fraction $p_{0,0}$ by subtracting the total loss fraction
from $1$:

	\begin{equation}
	\label{eq:p00new}
p_{0,0} = 1 - 2p_{0,1} - 2p_{0,2} - 4p_{1,1} .
	\end{equation}

\noindent
Then we iterate the calculation for the other coefficients according
to Eq.~\ref{eq:p01p11}. Merely three iterations could make
the  coefficients convergent, and the sum of PSF practically unity
(see Fig.~\ref{fig:sumpsf}, and the maximal deviation is less than
$0.0001$). The convergent PSF coefficients, shown as filled circles in
Fig.~\ref{fig:psflin} and Fig.~\ref{fig:psfqua}, are almost identical
whether the linear or quadratic fitting is used to derive the Poisson variance
$V$. In addition, $p_{0,0}$ decreases linearly with the signal level, while the
others increase linearly with the signal level. For instance, at the maximal
signal level of $6 \times 10^4$ ADU, about $9\%$ electrons in a pixel
will diffuse out and be captured by the neighbor pixels: $2\%$ by each
nearest vertical neighbor and $0.8\%$ by each second nearest vertical
neighbor and each diagonal neighbor. These results demonstrate that
this charge-sharing PSF model is simple but effective to explain the
nonlinear PTC.

On the other hand, the convergent $p_{0,0}$ differs a little from the
initial $p_{0,0}$ for the linear fitting (Fig.~\ref{fig:psflin}), 
and the subsequent coefficients do not vary much, either.  
However, in the case of the quadratic fitting (Fig.~\ref{fig:psfqua}), 
the slope of the convergent $p_{0,0} - signal$
relationship becomes less steep than the initial $p_{0,0}$,
thus the subsequent coefficients vary a little more
for high signal levels. This implies that the quadratic fitting
overestimates the Poisson noise, and consequently underestimates the
gain. Conversely, the results of $p_{0,0}$ could be used to estimate
the Poisson noise and the ``gain'' parameter more precisely.

\section{EFFECTS ON PHOTOMETRY}
\label{sec:effect}

\begin{figure}
   \begin{center}
   \begin{tabular}{l}
   \includegraphics[height=6cm]{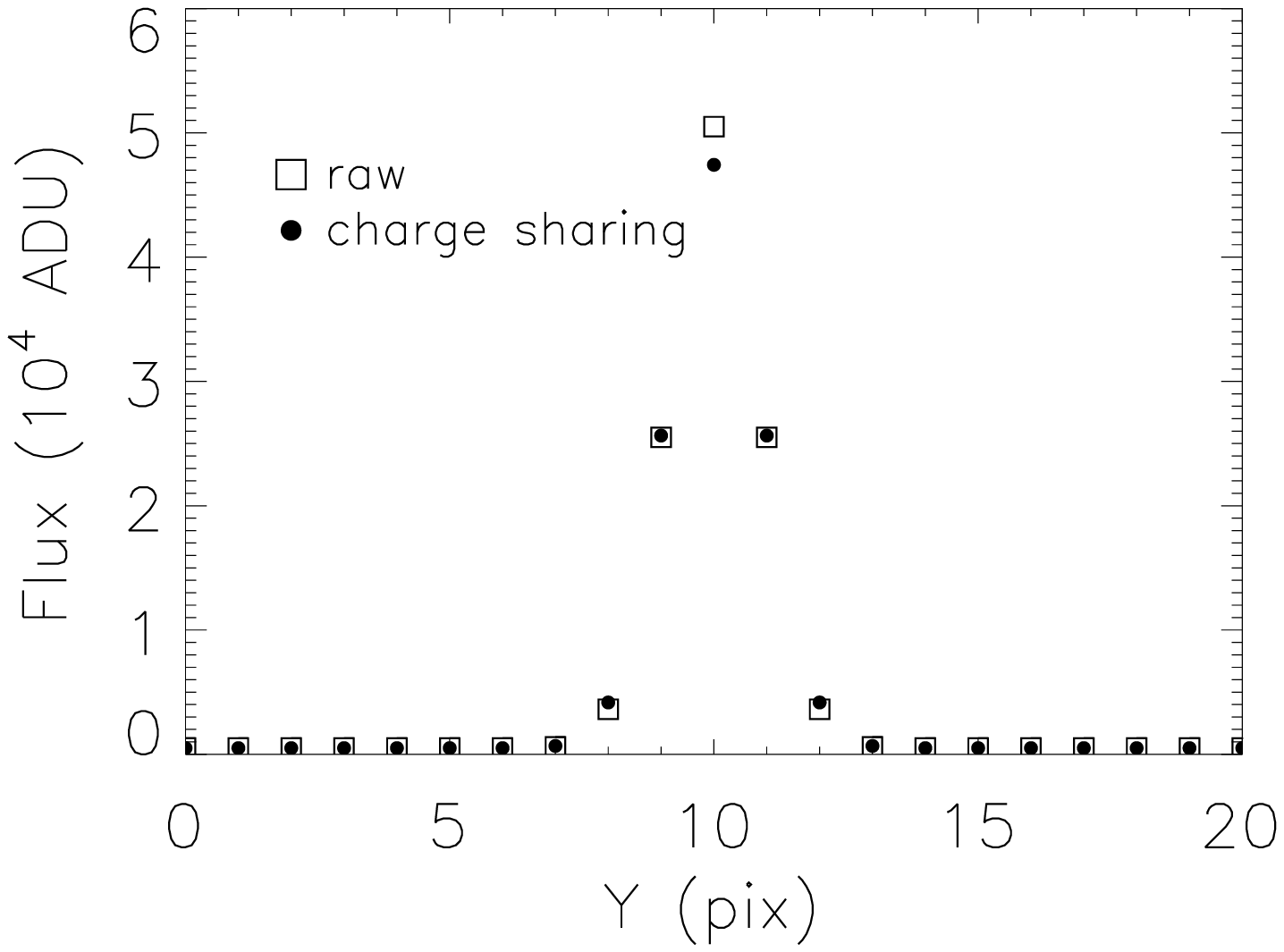}
   \includegraphics[height=6cm]{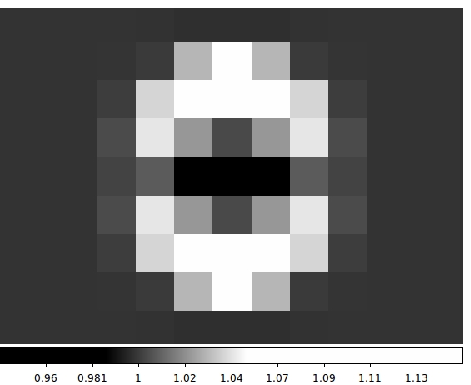}
   \end{tabular}
   \end{center}
   \caption[profile]
   {\label{fig:profile} How a Gaussian profile is changed by the charge sharing PSF. Left: the profile along Y-axis, the open squares are the raw profile, and the filled circles are the convolved profile. Right: the ratio of pixel values between raw image and convolved image.}
   \end{figure}

The impact of the charge sharing effects on photometry needs to be
investigated.
Firstly, the shape of a star will be changed due to this extra PSF.
For example, in Fig.~\ref{fig:profile} the open squares illustrates a
star with a Gaussian profile with a Full Width at Half Maximum (FWHM) of 2
pixels and a peak flux of $5 \times 10^4$ ADU. Convolved with the 
charge-sharing PSF derived in Sec.~\ref{sec:model}, the shape is shown by
filled circles in Fig.~\ref{fig:profile}. The resulted shape is more
complicate than being just broadened, because the charge-sharing PSF is dependent
on signal level. A brighter pixel will lose a larger fraction of electrons,
and the peak pixel will lose electrons with the largest
fraction. So the final shape is not exactly Gaussian any more, but
instead with the peak slightly sliced off. In Fig.~\ref{fig:profile},
the left figure shows how the profile along Y-axis changes. Moreover,
the charge sharing effect is much stronger along Y-axis than along X-axis. As
a result, the shape tends to be fatter along Y-axis, i.e. the shape is
elongated. The ratios of pixel values between raw image and the convolved
image (Fig.~\ref{fig:profile}, right) show that the central pixels are
abated, while their neighbors along Y-axis are obviously enhanced. 

\begin{figure}
   \begin{center}
   \begin{tabular}{c}
   \includegraphics[height=7cm]{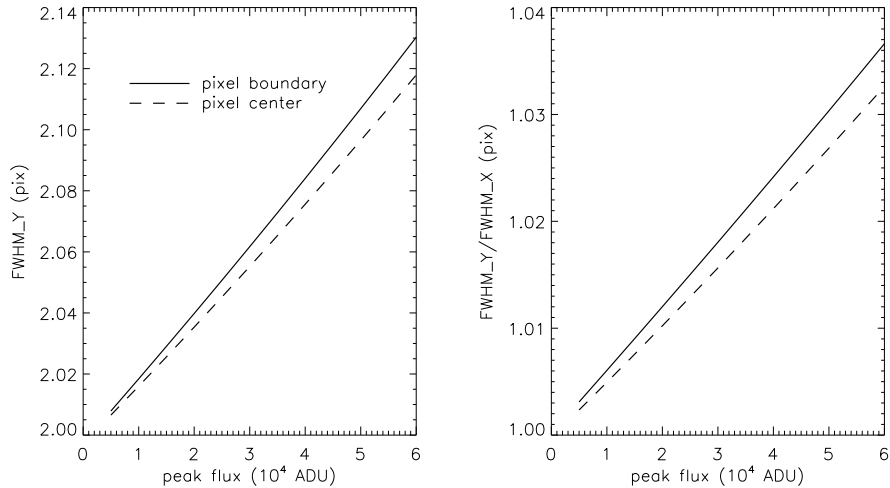}
   \end{tabular}
   \end{center}
   \caption[psf_flux]
   {\label{fig:psf_flux} The signal dependent shape: FWHM along Y-axis
(left), and the elongation (right). The stars are located at the
boundary between two pixels (solid lines), or at the center of a pixel
(dashed lines).}
   \end{figure}

In order to show how the FWHM changes with the peak flux, we simply
fit the convolved profiles with a Gaussian profile.
As seen in Fig.~\ref{fig:psf_flux}, 
the FWHM of a star increases linearly with its peak value.
For the brightest stars with a peak value of
$6 \times 10^4$ ADU, their FWHMs are about $0.12$ pixel larger than
those of the faintest stars, while their shapes are elongated by a
factor of more than $3\%$. 

However the FWHM dependence on peak flux varies with the intra-pixel
location of a star's centroid.  
As the centroid shifts inside a pixel, the pixels surrounding the peak
pixel will have different signal levels, corresponding to various
charge-sharing PSFs. The final profile will consequently vary a little.
Fig.~\ref{fig:psf_flux} illustrates two extremes with simulations: 
with the star centroid at the pixel boundary and center, respectively. 
The variations between the two extremes in FWHM
and elongation also increase with the peak flux, and their maximums are
about $10\%$. Although small, this second-order effect may be
considered for
applications requiring extreme precision on shape measurements.

The flux dependent shape will bias the PSF photometry when the measured
star has a different flux from the stars used to derive the PSF.
However, the aperture photometry would be less affected, because the
charge-sharing distance is essentially short. For bright stars,
large apertures are often adopted to improve the signal-to-noise ratio
(SNR), in which the charge loss fractions are essentially small. On
the other hand, for faint stars, although small apertures are
adopted to reduce the noise from background, the intrinsic Poisson
noise is still much larger than the bias caused by charge sharing,
which is also very weak at low signal levels. 
One complication would be to carefully define apertures of different
sizes for stars of different brightness.
The existence of the charge-sharing PSF can affect the photometry
using the difference image analysis (DIA, i.e., image subtraction)
dramatically, if only space-varying PSF is considered.  To achieve high precision, 
the charge-sharing PSF should also be taken into account to derive the convolve kernel.
Besides, the flat frames to correct the variations in the pixel-to-pixel 
sensitivity wil be influenced, because the charge sharing smoothes these 
variations.  It is more critical for dome flat and twilight flat, which are 
usually obtained at high signal level, while it is insignificant for super-sky 
flat, whose signal levels are essentially low.

\section{CONCLUSIONS}
\label{sec:conclusion}

We have investigated the PTC of AST3 CCD, and found that it becomes nonlinear at
about $1/3$ full well. While the linear fitting to PTC only suitable for lower
signal range, the quadratic function fits the entire signal range well.
However, the slopes of the linear fitting and the linear term of the quadratic
fitting vary by about $10\%$, resulting in $gain = 1.64$, $1.44$
$e^-$/ADU, respectively. The sub-Poisson noise suggests correlations
between neighboring pixels. We derive the 2D correlation coefficients,
and find the correlations increase with the signal level, decay rapidly with
the separation (distance), and are larger along Y-axis (CCD columns) than along
X-axis.  However, the nearest horizontal neighbors exhibit nearly negligible correlation,
whereas the second nearest horizontal neighbors show slight
correlation.  We do not have an explanation for this phenomena, yet.  

The correlation has been proposed to be caused by charge sharing when
the Coulomb forces produced by collected charges in a pixel reduce its
potential well and repel the subsequent charges. We present a signal
dependent charge-sharing PSF model to describe what fractions of
charges generated in one pixel are shared with its neighbors. The
charge-sharing PSF $p_{i,j}$ denotes the fraction of charges generated
in one pixel but finally collected by its neighbor pixel $(i,j)$. According to
the correlation results, we approximately take into account $p_{i,j}$
with significant correlations, which are $p_{0,\pm 1}$, $p_{0,\pm 2}$,
$p_{\pm 1,\pm 1}$ as well as $p_{0,0}$ which is the remained fraction
of charges in the original sharing pixel.
Then the nonlinearity of PTC and the correlations can be expressed in
terms of $p_{i,j}$. To solve $p_{i,j}$ conveniently, some reasonable
approximations are used based on the fact that $p_{0,0} \gg p_{0,1} \sim
p_{0,2} \sim p_{1,1}$. Although the linear fitting and the quadratic fitting
to the PTC
infer different nonlinear level, three iterations successfully
make $p_{i,j}$ convergent and consistent. Therefore the convergent $p_{0,0}$ is
supposed to derive a more precise gain, which will be investigated in
the future.

We demonstrate the shared fraction of charges increases linearly with
the signal level,
while the remained fraction decreases linearly with the signal level. This
signal dependent charge-sharing PSF will alter stars' profiles to be fatter for
brighter stars. Besides, the sharing fractions are larger along Y-axis
than X-axis, hence the stars' profiles will be elongated. Therefore
the PSF photometry will be biased if the measured star has a different
flux from the stars used to derive the PSF.  It is the same for image
subtraction photometry which largely depends on the PSF models.
However, we argue that the
aperture photometry is less influenced because the charge sharing decay rapidly 
with the separation distance. In addition, the dome flat and twilight flat are usually
taken at high signal levle, therefore will be smoothed by the charge sharing, and should 
be exposed at appropriate siganl level according to their CCD's charge sharing degree.

Our model is still simple with a relatively small charge-sharing PSF.
In the future, we will accurately measure the stars' shape in the
images from AST3, by removing other instrumental effects such as the
telescope tracking error. Then we could compare our
model with other models. Ours can be regarded as the integral
effect of the model in Ref.\,\citenum{Antilogus2014}, nevertheless herein
we only consider the signal level of one pixel, without those of its
neighbors. In the flat frames, the signal levels in individual pixels are
almost the same, and the differences are caused by Poisson noise. In
the scientific frames, however, the signal levels vary quite a lot
between stars and the sky background. Even for the stars, the signals drop
dramatically with the distance from the center. Take all these into
account, we will improve our model to be more accurate. In addition, a
broader PSF will be considered to include farther neighbors, which
also exhibit slight correlations.

\acknowledgments

This work has been supported by the National Basic Research Program of
China (973 Program) under grand No. 2013CB834900, the Chinese Polar
Environment Comprehensive Investigation \& Assessment Programmes under
grand No. CHINARE2014-02-03, and the National Natural Science
Foundation of China under grant No. 11003027, 11203039, and 11273019.


\bibliography{ref} 

\begin{thebibliography}{1}

\bibitem{Janesick2000}
Janesick, J.~R.,  [{\em Scientific Charge-Coupled
  Devices}{\nolinebreak\hspace{0.1em}]}, SPIE Press monograph, Washington
  (2001).

\bibitem{Downing2006}
{Downing}, M., {Baade}, D., {Sinclaire}, P., {Deiries}, S., and {Christen}, F.,
  ``{CCD riddle: a) signal vs time: linear; b) signal vs variance:
  non-linear},'' {\em Proc. SPIE} {\bf 6276},  627609 (2006).

\bibitem{Downing2009}
Downing, M., B. D. D.~S. and Jorden, P., ``{Bulk silicon CCDs, Point Spread
  Function and Photon Transfer Curves: CCD testing activities at ESO},'' {\em
  Presented at Detectors for Astronomy Workshop} (2009).

\bibitem{Downing2013}
Downing, M. and Sinclaire, P., ``The ccd riddle revisted: Signal versus time
  – linear signal versus variance – non-linear,'' {\em Proc. SDW} (2013).

\bibitem{Astier2013}
{Astier}, P., {El Hage}, P., {Guy}, J., {Hardin}, D., {Betoule}, M., {Fabbro},
  S., {Fourmanoit}, N., {Pain}, R., and {Regnault}, N., ``{Photometry of
  supernovae in an image series: methods and application to the SuperNova
  Legacy Survey (SNLS)},'' {\em {A$\&$A}}~{\bf 557},  55 (2013).

\bibitem{Stefanov2014}
Stefanov, K., ``{A Statistical Model for Signal-Dependent Charge Sharing in
  Image Sensors},'' {\em IEEE Transactions on Electron Devices}~{\bf 61(1)},
  110--115 (2014).

\bibitem{Antilogus2014}
{Antilogus}, P., {Astier}, P., {Doherty}, P., {Guyonnet}, A., and {Regnault},
  N., ``{The brighter-fatter effect and pixel correlations in CCD sensors},''
  {\em Journal of Instrumentation}~{\bf 9},  C3048 (2014).

\bibitem{Cui2008}
Cui, X., Yuan, X., and Gong, X., ``{Antarctic Schmidt Telescopes (AST3) for
  Dome A},'' {\em Proc. SPIE}~{\bf 7012},  70122D (2008).

\bibitem{Ma2012}
Ma, B., Shang, Z., Wang, L., Boggs, K., Hu, Y., Liu, Q., Song, Q., and Xue, S.,
  ``{The test of the 10k x 10k CCD for Antarctic Survey Telescopes (AST3)},''
  {\em Proc. SPIE} {\bf 8446},  84466R (2012).

\end{thebibliography}
\bibliographystyle{spiebib}
\end{document}